\documentclass[lettersize,journal]{IEEEtran}
\usepackage{amsmath,amsfonts}
\usepackage{algorithmic}
\usepackage{array}
\usepackage[caption=false,font=normalsize,labelfont=sf,textfont=sf]{subfig}
\usepackage{textcomp}
\usepackage{stfloats}
\usepackage{url}
\usepackage{booktabs}
\usepackage[noadjust]{cite}
\usepackage{verbatim}
\usepackage{graphicx}
\usepackage{color}
\usepackage{soul}

\def\BibTeX{{\rm B\kern-.05em{\sc i\kern-.025em b}\kern-.08em
    T\kern-.1667em\lower.7ex\hbox{E}\kern-.125emX}}
\usepackage{balance}
\begin{document}
\title{Reconfigurable Intelligent Surface for Physical Layer Security in 6G-IoT: Designs, Issues, and Advances}
\author{Waqas Khalid, M. Atif Ur Rehman, Trinh Van Chien, Zeeshan Kaleem, Howon Lee, Heejung Yu
\thanks{Waqas Khalid is with the Institute of Industrial Technology, Korea University, Sejong 30019, Korea (email:waqas283@\{gmail.com, korea.ac.kr\}).

M. Atif Ur Rehman is with the Manchester Metropolitan University, Manchester, UK (email:m.atif.ur.rehman@mmu.ac.uk).

Trinh Van Chien is with the Hanoi University of Science and Technology, Hanoi, Vietnam (email:chientv@soict.hust.edu.vn).

Zeeshan Kaleem is with the COMSATS University
Islamabad, Wah Campus, Pakistan (email:zeeshankaleem@gmail.com) 

Howon Lee (Corresponding author) is with the School of Electronic and Electrical Engineering and IITC, Hankyong National University, Anseong 17579, South Korea (e-mail: hwlee@hknu.ac.kr).

Heejung Yu (Corresponding author) is with the Department of Electronics and Information Engineering, Korea University, Sejong 30019, Korea (email:heejungyu@korea.ac.kr).}
}

\markboth{IEEE Internet of Things Journal,~Vol.~x, No.~x, xx}%
{How to Use the IEEEtran \LaTeX \ Templates}

\maketitle

\begin{abstract}
Sixth-generation (6G) networks pose substantial security risks because confidential information is transmitted over wireless channels with a broadcast nature, and various attack vectors emerge. Physical layer security (PLS) exploits the dynamic characteristics of wireless environments to provide secure communications, while reconfigurable intelligent surfaces (RISs) can facilitate PLS by controlling wireless transmissions. With RIS-aided PLS, a lightweight security solution can be designed for low-end Internet of Things (IoT) devices, depending on the design scenario and communication objective. This article discusses RIS-aided PLS designs for 6G-IoT networks against eavesdropping and jamming attacks. The theoretical background and literature review of RIS-aided PLS are discussed, and design solutions  related to resource allocation, beamforming, artificial noise, and cooperative communication are presented. We provide simulation results to show the effectiveness of RIS in terms of PLS. In addition, we examine the research issues and possible solutions for RIS modeling, channel modeling and estimation, optimization, and machine learning. Finally, we discuss recent advances, including STAR-RIS and malicious RIS.
\end{abstract}


\begin{figure}[t!]
\centering
\includegraphics[width=3.25in]{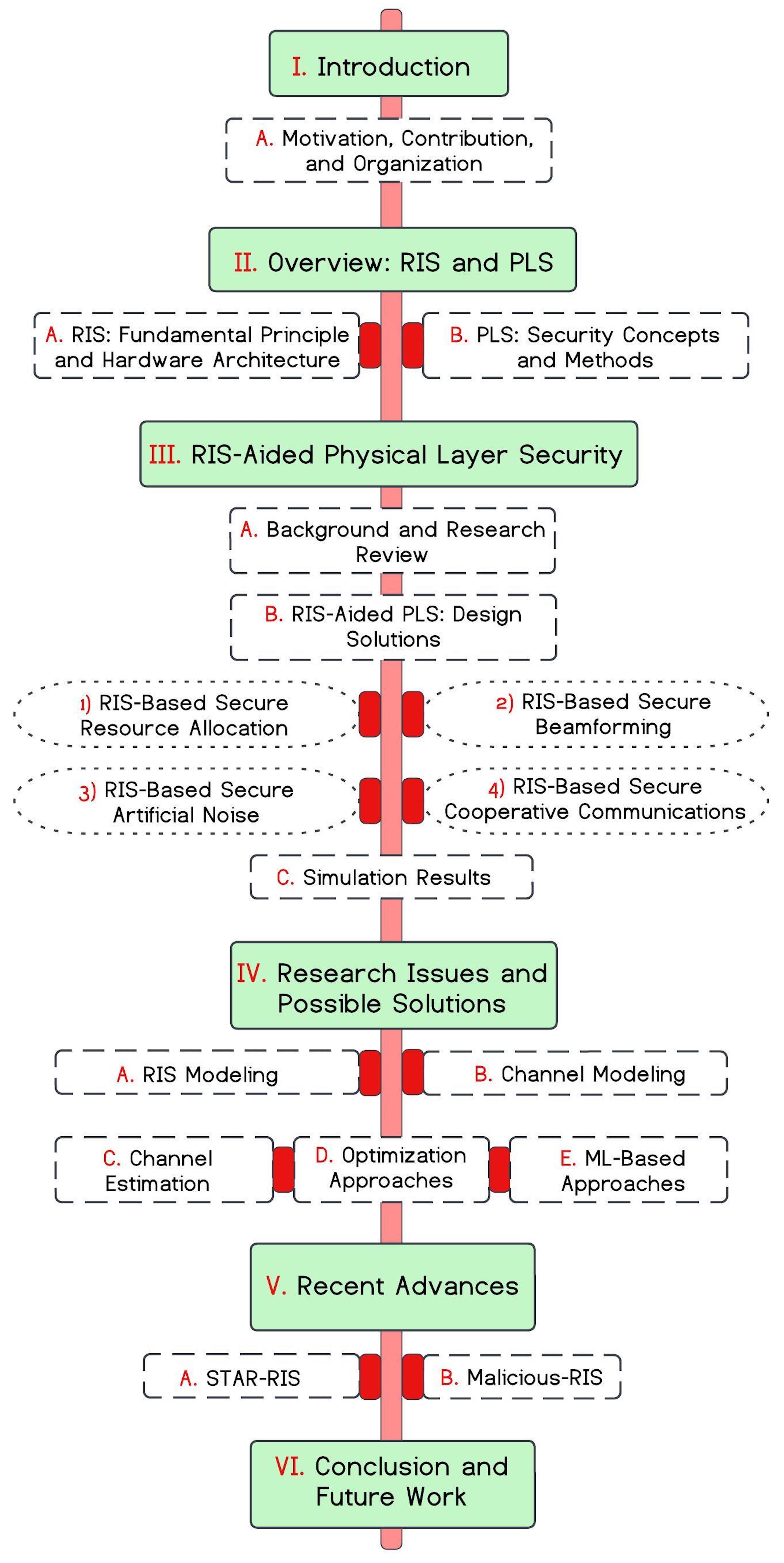}
\caption{Organization of the paper.}
\label{fig1}
\end{figure}

\section{Introduction}
\label{section1}
Sixth-generation (6G) networks offer innovative use cases, unprecedented services, and revolutionary applications owing to their high transmission rates, significantly low latency, and enhanced reliability. The Internet of Things (IoT) is a recently developed communication paradigm that provides all-encompassing connectivity solutions in 6G \cite{m1}. Nevertheless, 6G-IoT networks pose severe security risks due to the broadcast nature of wireless channels, large amounts of confidential and privacy-sensitive data (e.g., online banking, personal tracking, home control, health monitoring, and industrial automation), and an exponential increase in attack vectors. A 6G-IoT network must be hyper-secure from the physical to the application layers. In this regard, physical layer security (PLS) is an effective method of securing legitimate transmissions by leveraging the dynamic nature of wireless {communications \cite{9367233}}.

Millimeter-wave (mmWave) and sub-terahertz (sub-THz) frequencies offer considerable advantages despite unique propagation challenges, such as high sensitivity to blockages and severe penetration loss \cite{m1,9367233}. As an example, point-to-point links at these frequencies have significantly higher transmission losses than those at sub-6 GHz systems. Therefore, future algorithms and protocols must be designed to mitigate the adverse effects of an uncontrolled radio environment. A conventional transmission strategy optimizes a transmitter (Tx) and receiver (Rx) through various approaches, such as multiple antennas, complex signal processing algorithms, sophisticated encoding and decoding procedures, and advanced communication protocols. However, the Tx and Rx design cannot be extremely complex in certain application scenarios, such as resource-constrained IoT and industrial field devices.

{Table \ref{table1} lists the entities discussed in the existing literature for controlling the radio environment, including reconfigurable intelligent surfaces (RISs) \cite{m2}, passive analog repeaters \cite{m2a,m2b}, decode-and-forward (DF) relays \cite{m2c}, and amplify-and-forward (AF) relays \cite{m2d}. Particularly, RIS is an emerging technology that can provide a dynamic, controllable, and programmable radio environment}. Further, different metasurface implementations have been developed for the full space environment, including simultaneous transmitting and reflecting-RIS (STAR-RIS) and intelligent omni-surface (IOS) \cite{m3}. Owing to the dependence of the PLS on a richly scattered environment, the dynamic control of channels achieved by the RIS can significantly improve security in 6G-IoT networks.

\begin{table*}[t!]
\caption{Entities that control the radio environment }
\centering
%% \tablesize{} %% You can specify the fontsize here, e.g., \tablesize{\footnotesize}. If commented out \small will be used.

\begin{tabular}{ |p{1.4cm}| p{1.6cm}|  p{5.8cm}| p{7.4cm}|}
\toprule
\textbf{{Entity}}& \textbf{{Type}} & \textbf{{Advantages}} & \textbf{{Disadvantages}}\\
\midrule
RIS
&
Transparent
&
\begin{itemize}
\item 	Full-duplex (FD) operation
\item Low power consumption 
\item 	Noise is not created due to the absence of active components (introduces a phase shift)
\item 	Simple RF processing
\end{itemize}
&
\begin{itemize}
\item 	Must be controlled by a base station (BS) with an additional control link (wired or wireless) 
\item Discrete phase shift and phase noise in the hardware
\item Complex channel estimation and reflection pattern design
\end{itemize} \\ 
\midrule
Passive analog repeater
&
Transparent
&
\begin{itemize}
\item 	FD operation
\item 	Simple and cost-effective design
\item 	Not require any control from a BS 
\item 	Fully passive (only reflection)
\end{itemize}
&
\begin{itemize}
\item 	No flexibility, fixed beams
\item	Reflects the signal without amplification (limited achievable transmission range)
\item	Dependence on signal strength
\end{itemize} \\
\midrule
{DF}-relay
&
Regenerative
&
\begin{itemize}
\item 	Supports half-duplex (HD) and FD modes
\item 	No noise amplification problem
\item 	Signals received can be intentionally analyzed and decoded
\end{itemize}
&
\begin{itemize}
\item	Higher hardware complexity
\item	High power consumption compared to RIS
\item	Signal processing operations are more complex than RIS
\end{itemize} \\
\midrule
{AF}-relay
&
Transparent
&
\begin{itemize}
\item	Supports both HD and FD modes
\item	Signal is boosted (amplified) and forwarded without further processing
\end{itemize}
&
\begin{itemize}
\item	Noise amplification
\item	Hardware complexity lower than DF
\item	High power consumption compared to RIS
\item	Signal processing operations are more complex than RIS
\end{itemize} \\
\bottomrule
\end{tabular}
\label{table1}
\end{table*}

{Continuing next, the motivation, contribution, and organization of the paper will be presented as follows:}

\textit{Motivation}: PLS can be used independently or in combination with post-quantum cryptography for the security of simple 6G-IoT devices \cite{nt1}. PLS methods may be ineffective in unfavorable propagation conditions because they rely on noise and fading variations in wireless channels. Security can be enhanced in 6G-IoT networks with dynamically controlled channels provided by RIS \cite{nt2}. For example, an RIS can completely exploit PLS advantages caused by channel propagation characteristics, spatial diversity, beamforming, and cooperative communications. RIS-aided PLS designs can be applied to both 5G and 6G without distinction \cite{nt3}, however, they require precise configuration, placement of the RIS, strict hardware requirements, and elements coupling. In addition, RIS-aided PLS solutions can be selected based on the design scenario and communication objective, balancing security performance and implementation complexity \cite{nt1,nt2,nt3}. 

\textit{Contribution}: Based on the aforementioned motivation, we present a general framework for designing and implementing PLS solutions using RIS to prevent eavesdropping and jamming attacks in the 6G-IoT wireless paradigm. The RIS is introduced in terms of fundamental principles and hardware architecture, and the PLS is presented in terms of background concepts, leveraging sources, and security methods. We discuss the theoretical background and literature review of RIS-aided PLS and present design solutions related to resource allocation, beamforming, artificial noise (AN), and cooperative communications. Simulation results are also provided to verify the effectiveness of RIS in terms of the PLS performance measures, e.g., secrecy rate and secrecy outage probability, under various network configurations. In addition, we examine the research issues and possible solutions for RIS modeling, channel modeling, channel estimation, optimization, and machine learning (ML), and discuss recent advances, such as STAR-RIS and malicious RIS.

\textit{Organization}: Fig. \ref{fig1} illustrates the organization of this article. Section \ref{section2} introduces RIS and PLS. Section \ref{section3} presents the background, literature review, and design solutions of the RIS-aided PLS, along with simulation results. We discuss research issues and possible solutions, and recent advances, respectively, in Sections \ref{section4} and \ref{section5}. Finally, Section \ref{section6} concludes the article.

\section{Overview: RIS and PLS}
\label{section2}
In this section, an RIS is introduced in terms of fundamental principles and hardware architecture, and PLS is discussed in terms of background concepts, leveraging sources for security, and security requirements and methods.

\subsection{RIS: Fundamental Principle and Hardware Architecture}
\begin{figure}[t!]
\centering
\includegraphics[width=3.4in]{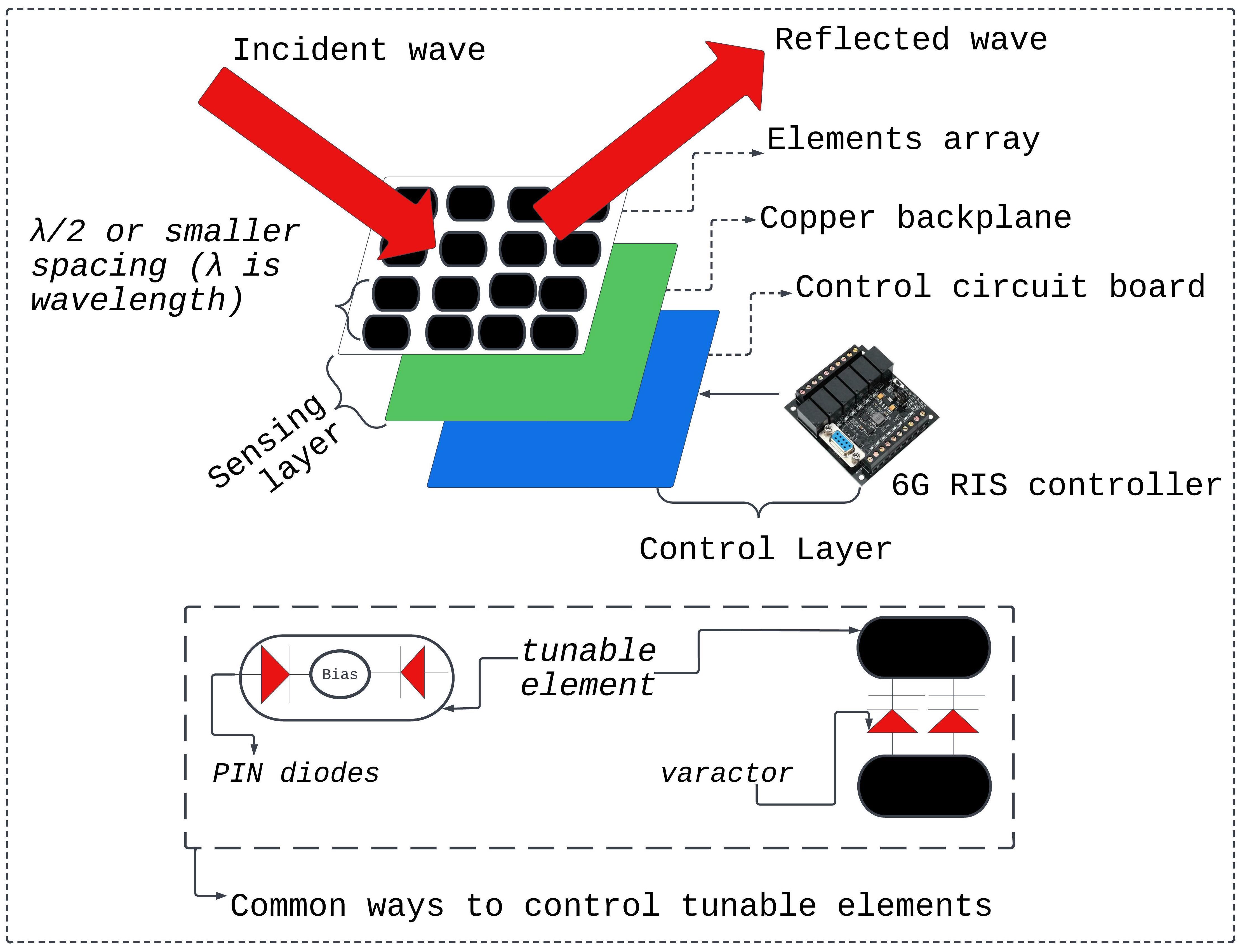}
\caption{Schematic of an RIS, where a 6G controller configures elements to control the reflection depending on PLS requirements.}
\label{fig2}
\end{figure}

RIS is a hardware transmission technology that reconfigures wireless channels and establishes a smart radio environment. As shown in Fig. \ref{fig2}, the RIS configuration comprises three layers and a controller with the following details \cite{nt4}:
\begin{itemize}
\item The outer layer has flexible, discrete, sub-wavelength-sized elements. An effective and scalable method of modifying tunable elements and controlling electromagnetic (EM) reflection is to use a diode array, such as a positive-intrinsic-negative (PIN) or varactor diode array. A voltage setting can alter the state of the diode, thereby controlling the tunable elements.
\item A copper backplane is placed in the middle layer to prevent signal leakage.
\item In the inner layer, a control circuit board adjusts the reflection coefficients (amplitude and phase). It connects the RIS controller to the reflecting elements, i.e., it receives and transmits voltage signals.
\item An RIS controller adjusts the phase of the incident signal based on commands from the central controller. It is implemented using a field-programmable gate array (FPGA), which acts as a gateway between other network components (e.g., BSs and user terminals).
\end{itemize}

By manipulating the incident EM waves, RIS can produce the desired scattering and reflection patterns while providing solutions to radio frequency (RF) impairments and signal distortions. Through controlled scattering and multi-path components, RIS extends transmission coverage under harsh propagation conditions, such as blockage and shadowing, and ensures wireless connectivity over multiple concurrent links \cite{m4}. Accordingly, better coverage and reliability can be achieved without sophisticated signal processing or RF operations. The radio environment is not considered an uncontrollable and random entity because it is part of the design parameter in an  optimization space. Joint optimization of the Tx, Rx, and wireless channels can significantly improve performance metrics, including rate, latency, reliability, privacy, energy efficiency, and massive connectivity. Through intelligent placement and effective beamforming designs, the RISs can enable advanced wireless functions, such as a signal boost in mmWave and sub-THz bands, power consumption reduction, increased throughput, interference suppression, and reliable and secure reception. Hence, 6G-IoT networks can maximize their potential \cite{nt4,m4}.

\subsection{PLS: Security Concepts and Methods}
\subsubsection{Background Concepts}
Cryptography and PLS are promising solutions to secure wireless transmissions. Cryptography provides security at the higher layers using classical encryption-based algorithms and protocols, and PLS manipulates the intrinsic characteristics of the wireless channels \cite{m5}. Cryptography-based solutions have the following key features:
\begin{itemize}
\item In general, cryptography and encryption are complex and vulnerable.
\item Cryptography and encryption (e.g., symmetric and asymmetric) require complicated operations (resources and capabilities) at Tx and Rx. Security can be compromised by inexhaustible computing power, network latency, and network overhead.
\item Hierarchical decentralized architectures, delay-sensitive and power-limited applications, and large-scale and highly mobile systems are unsuitable.
\end{itemize}

In contrast, PLS-based solutions are characterized by the following key features:
\begin{itemize}
\item 	PLS can complement or replace cryptography. Secure transmission can be achieved through signal design and signal processing (no key management, exchange, or maintenance required).
\item 	PLS solutions (e.g., beamforming, AN, and resource allocation) may not require computational operations in the end users.
\item PLS is simple, flexible, scalable, and lightweight, reducing complexity and overhead. It offers security solutions that are cost-effective while balancing implementation complexity.
\item A suitable solution for applications with power- and delay-sensitive requirements, as well as evolving physical layers and networking infrastructures (where security is implemented at the edge).
\end{itemize}
 
As each type of secure solution has advantages and disadvantages, selecting a specific approach will depend on the security requirements, type of security attack, and type of communication system. A combined approach can often provide an advanced security level \cite{8302922}. From an information-theoretic perspective, PLS has garnered considerable interest in industrial, academic, and research communities over the last few decades, providing unbreakable, proven, and quantifiable security \cite{YU2020611}.

\subsubsection{Leveraging Sources for Security}
Random and time-varying wireless channels with limited coherence distance, time, and bandwidth present challenging propagation environments. Nevertheless, independent channels with distinct observable parameters are crucial for data security. Additionally, a wireless link can be authenticated or secured using the unique characteristics of the RF front end and the physical properties of the radio environment. For example, hardware fingerprints, such as phase noise, carrier frequency offset, in-phase/quadrature imbalance, non-linearity of the power amplifier, and antenna imperfections, can complement channel-based PLS methods \cite{m6a}. Particularly, channel-based PLS methods are more appropriate for relatively stationary and indoor environments (e.g., IoT). In contrast, RF-based PLS methods provide more stable and secure solutions in rapidly changing environments. Neither of these types of PLS introduces unnecessary overhead, as channel estimation is an integral part of wireless communications, as well as measuring the imperfections in the RF transceivers to ensure reliable communications. Moreover, physical measurements such as distance, speed, angle, size, and constituent materials can be incorporated into environment-based PLS methods \cite{m6}. In this article, we focus on channel-based PLS solutions to address the security of 6G-IoT networks by leveraging the propagation characteristics of wireless channels with RIS.

\subsubsection{Security Requirements and Malicious Attacks}
\begin{figure}[t!]
\centering
\includegraphics[width=2.6in]{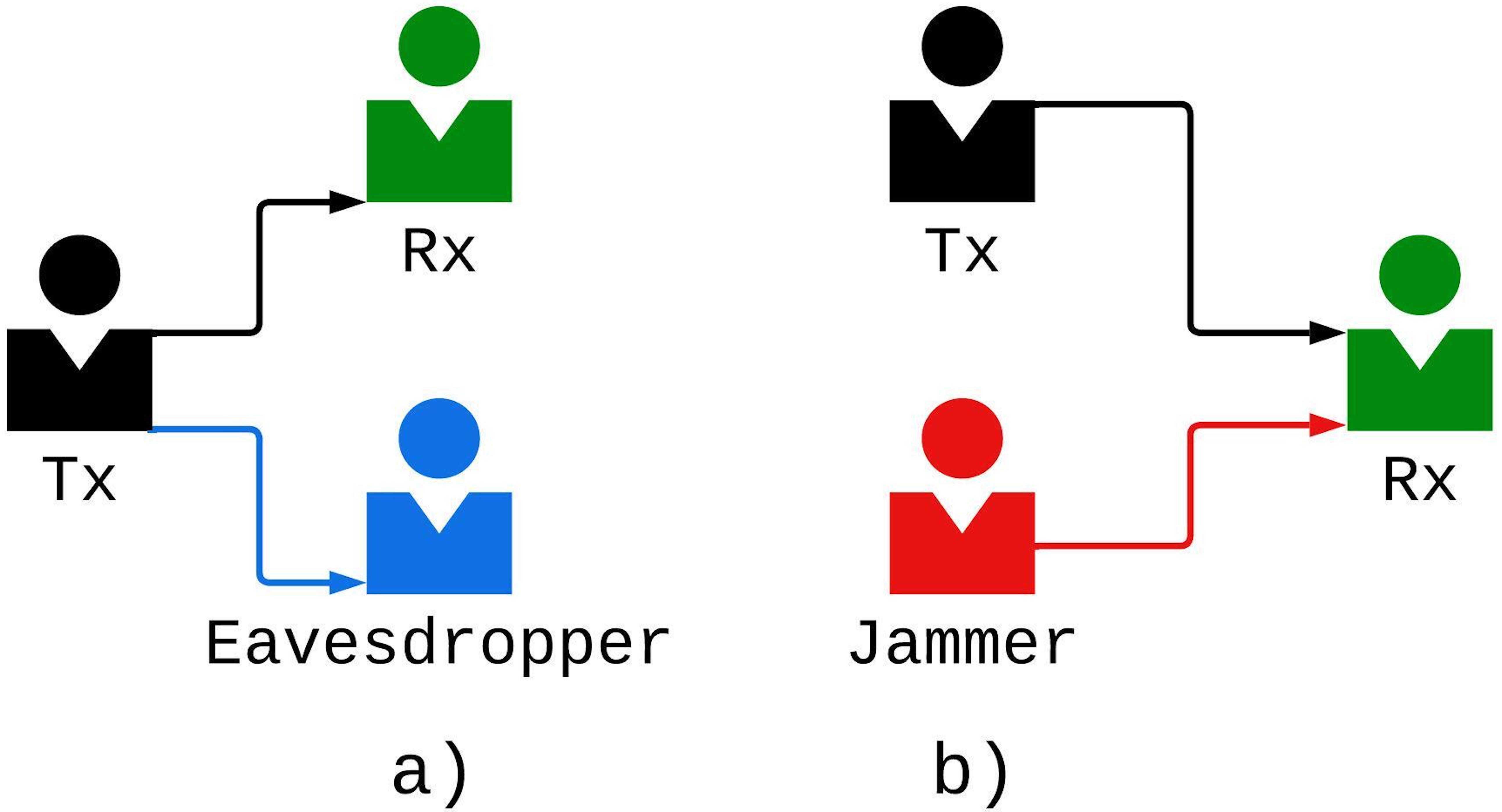}
\caption{Physical layer attacks: a) Eavesdropping attacks; b) Jamming attacks.}
\label{fig3}
\end{figure}

A top-level taxonomy of confidentiality and availability can be used to describe the requirements for secure communication systems. Confidentiality refers to exposing information to Rx without being detected by an eavesdropper. Availability refers to the ability of the Tx to access wireless networks at any time and from anywhere. Owing to the broadcast and superposition properties of wireless networks, wireless transmissions are susceptible to various security breaches, including passive eavesdropping attacks (intercepting data) and active jamming attacks (disrupting transmissions) \cite{m7}. Fig. \ref{fig3} illustrates certain examples of eavesdropping and jamming attacks in which the eavesdropper overhears the information without taking any proactive measures while the jammer injects an interfering signal to block legitimate transmissions. The transmission of high-power noise or false data by a jammer result in active malfunctions, in which the Tx cannot access the channel, or the Rx receives distorted information.

\subsubsection{Security Methods}
Anti-jamming solutions make it difficult for jammers to disrupt legitimate transmissions. Frequency hopping involves switching between different frequencies in a predetermined pattern (or pseudo-random code-based) to make the communications orthogonal to the jamming signal; a jammer is required to hop between frequencies to track the signal. Spread spectrum spreads the signal over a wide range of frequencies. Antenna diversity requires multiple antennas at the Rx to improve signal reception. Redundancy refers to transmitting the same information multiple times over different channels or utilizing error-correcting codes to recover the transmitted information. Physical separation involves separating the Rx by a large distance. Moreover, ML algorithms can be trained to detect and classify jamming signals and adapt transmission parameters in real-time to minimize jamming impact under varying conditions, thereby making anti-jamming communications more robust, flexible, and efficient. Generally, frequency hopping and spread spectrum require additional spectral resources and are ineffective owing to the diversity and dynamic nature of intelligent jammers. Furthermore, redundancy, antenna diversity, and physical separation may not be feasible. By contrast, anti-eavesdropping communications intentionally impair the reception of information at the eavesdropper by using techniques such as beamforming, interference cancellation, AN, resource control, directional antennas, and cooperative communications \cite{m8}. In general, multi-antenna beamforming involves high hardware costs, computational complexity, and energy consumption due to the massive RF chains and complex signal processing requirements. AN and cooperative communications present significant challenges related to deployment scalability, hardware costs, and power consumption. In PLS methods, the inherent randomness of the noise and fading variations in a wireless channel are exploited. Because of this, they are less effective in unfavorable propagation environments, which can be avoided by using RIS \cite{m9}.

\section{RIS-Aided Physical Layer Security}
\label{section3}
This section presents insights into RIS-aided PLS in the 6G-IoT wireless paradigm and discusses the theoretical background, literature overview, design solutions against jamming and eavesdropping attacks, and simulation results.

\subsection{Background and Research Review}
The RIS has many unique characteristics, including high-precision beamforming, rapid deployment, limited control messages, real-time configuration, affordable complexity, full-band response, FD transmission, superior power gain (without imposing thermal noise), cost-effectiveness, and a high level of energy and spectral efficiency. The RIS offers considerable performance advantages and compatibility with PLS methods, making it an ideal candidate for implementing and enhancing PLS \cite{YADAV2022}. In the straightforward RIS-aided PLS design, the Tx communicates with the Rx over a vulnerable communication link, while the RIS is deployed at a strategic location to exploit the characteristics of the wireless channel. The RIS can protect against eavesdropping by limiting the information of the secret messages extracted by the eavesdropper by improving the signal-to-interference-plus-noise ratio (SINR) at the Rx (by mitigating the fading effect and maintaining a stable channel) and/or degrading the eavesdropper (by providing multi-path fading attenuation over the channel). The RIS can also protect against jamming by enhancing the SINR at the Rx to make the interfering signal less effective at disrupting communications. Therefore, the optimal adjustment of the deployment location, number of elements, and reflection coefficients for the RIS can protect the Rx from jamming and eavesdropping attacks.

In terms of secure transmissions, RIS setups offer numerous advantages and attractive features. The RIS-aided PLS solutions can be applied to key 6G technologies such as mmWave/sub-THz communications, visible light communications, vehicular communications, unmanned aerial vehicle (UAV) communications, integrated sensing and communication, industrial IoT, simultaneous wireless information and power transfer (SWIPT), non-orthogonal multiple access (NOMA), and device-to-device (D2D) communications \cite{nt1,nt2,nt3,YADAV2022,m11,m12,m13,m14,m15}. As presented in Table \ref{table3}, the improved PLS performance under controlled channel gain through the RIS has been demonstrated under several scenarios, systems models, methodologies, performance metrics, and optimization problems \cite{m11,m12,m13,m14,m15,m16,m17,m18,m19,m20}. Despite extensive research efforts, the study and evaluation of RIS-aided PLS in 6G-IoT networks are still at an early stage. The design and implementation of RIS-aided PLS under variable network topologies and novel application scenarios require further investigation. In particular, research is required to determine the capabilities and limitations of design solutions, and their effective implementations, inherent complexity, and additional control variables.

 \begin{table*}[t!] 
 {\scriptsize
{\caption{Recent efforts toward the realization of 6G RIS-aided PLS.}
\centering
%% \tablesize{} %% You can specify the fontsize here, e.g., \tablesize{\footnotesize}. If commented out \small will be used.
\begin{tabular}{|p{.3cm} | p{1.2cm}|   p{1.4cm}  p{2cm}  p{1.4cm} | p{2cm}| p{1.4cm}| p{1.4cm} | p{.7cm}  p{1.4cm}  p{.8cm}| }
\toprule

\textbf{Ref.} & 

\textbf{System model }  &

 \;\;\;\;\;\;

  \;\;\;\;\;\;
  
------------
  
PLS metric  &\textbf{Design  and  optimization}
 
-------------------

 Design variables &

 \;\;\;\;\;\;
 
 \;\;\;\;\;\;
 
 --------------
  
   Optimization methodology  &\textbf{Insight remarks} &\textbf{Security attack} & \textbf{RIS-aided PLS design} &

  \;\;\;\;\;\;

------- 
   
   Channel model & \textbf{Modeling}
  
  -------------- 
   
Perfect CSI: Rx, malicious users
  & \;\;\;\;\;\;

---------
   
 RIS impairments\\   \midrule 
  \;  \cite{m11} &  THz-MIMO  &   Secrecy rate  & Tx hybrid beamforming and RISs phase shift designs &  Iterative algorithms (Classical)& Improved security for the blockage-prone THz & Eavesdropping & Beamforming  &  Rayleigh & \;  \textbf{$\checkmark$}  \; \;\textbf{$\times$} & Ideal \\ \midrule
   \; \cite{m12} &  IOS-SWIPT IoT (SISO)& Secrecy rate &   UAV trajectory and transmit and jamming powers, and IOS phase shifts &Iterative  SCA (Classical) &   IOS provides secure communications in full space & Eavesdropping & Friendly jammer & Rician &\;  \textbf{$\checkmark$}  \; \;\textbf{$\checkmark$}  & Ideal  \\
 \midrule
   \; \cite{m13} &  Aerial-RIS (SISO)& Transmission rate &   Deployment and beamforming design for A-RIS &Alternating optimization (Classical) &   
Optimal ARIS design confirms jamming mitigation & Jamming & Passive beamforming & Rayleigh &\;  \textbf{$\checkmark$}  \; \;\textbf{$\checkmark$}  & Ideal  \\
 \midrule
  \; \cite{m14} &  Uplink SIMO & 
System rate&   Transmission energy of Tx and phase shifts of RIS & Deep Q network (ML) &   DQN algorithm can tackle the dynamic variations & Jamming & Passive beamforming & Rayleigh &\;  \textbf{$\checkmark$}  \; \;\textbf{$\checkmark$}  & Discrete phase shifts \\
 \midrule
  \; \cite{m15} & Multi-layer ITAN & 
 EE maximization problem &   User's received decoder, terrestrial and aerial precoder, and RIS reflection&  Block coordinate descent (Classical)  &   RIS-Tx can reduce cost, bring new DoF, and facilitate a large-scale array  &  jamming
and eavesdropping & Passive beamforming & Rayleigh & \;  \textbf{$\checkmark$}  \; \;\textbf{$\times$}  & Ideal\\
\midrule
  \; \cite{m16}  & RIS-based back-scatter& 
 Minimize SINR of eavesdropper &   Transmit power at Tx and phase shift design at RIS & 

Convex semi-definite program (Classical) & Novel approach  to modulate the received signal into jamming via RIS
&  Eavesdropping & Passive beamforming & Rayleigh & \;   \textbf{$\checkmark$}  (\textbf{$\times$} \textbf{$\checkmark$})  & Ideal \\
\midrule
  \; \cite{m17} & Multi-antenna BS and NOMA users & 
 Sum rate under QoS, RIS, and SIC constraints &   Tx beamforming with jamming and RIS reflection designs & 

Iterative alternating optimization  (Classical) & Secrecy transmission via artificial jamming
&  Eavesdropping & Beamforming and jamming & Rician & \;   \textbf{$\checkmark$}  \;  \textbf{$\times$}  & Ideal \\
\midrule
  \; \cite{m18}  & RIS-SISO network & 
 Secrecy outage probability  &   RIS reflection coefficients, path loss,  eavesdropper location distributions  & 

Closed-form analysis & Secure communications with uncertain eavesdropper distributions 
&  Eavesdropping & RIS beamforming & Rayleigh & \;   \textbf{$\checkmark$}  \;  \textbf{$\times$}  & Discrete phase shifts\\
\midrule
  \; \cite{m19} & RIS-SWIPT system & 
 Secrecy rate under EH and transmit power constraints  &   Tx transmit power, user power splitting factors, and
RIS phase shift matrix & 

AO-based, and ML-based & DL methods provide AO algorithm-based performance with
lesser computation time
&  Eavesdropping & RIS beamforming & Rayleigh and Rician & \;   \textbf{$\checkmark$}  \;  \textbf{$\checkmark$} & Ideal\\
\midrule
  \; \cite{m20}  & STAR-RIS-NOMA network & 
 Secrecy rate   &  AN precoding, STAR-RIS design, and users decoding order  & Iterative alternating optimization (Classical) & STAR-RIS-aided AN designs for the users in full space
&  Eavesdropping & Full space beamforming via STAR-RIS & Rayleigh & \;   \textbf{$\checkmark$}  \;  \textbf{$\checkmark$} & Ideal\\
\bottomrule
\end{tabular}
\label{table3}}
}

\end{table*}

\subsection{RIS-Aided PLS: Design Solutions}
In challenging scenarios, RIS-aided PLS solutions have demonstrated their effectiveness when compared with conventional MIMO methods \cite{m15,m16,m17,m18,m19,m20}. This may include situations where the Rx has a higher secrecy rate requirement, the eavesdropper/jammer contains more antennas (or has a better channel condition), or the eavesdropper (jammer) is located closer to the Tx. In such cases, PLS provisioning using large-scale antenna arrays may be ineffective and spatial beamforming may not provide sufficient secrecy gain. Furthermore, securing transmissions via combined methods, such as transmit beamforming with AN and/or cooperative methods, may also be challenging. Controlling the strength and direction of the reflected signal by using RIS and performing signal processing and optimization for the Tx and Rx presents novel security benefits. A properly configured RIS can optimize the radio environment under challenging propagation conditions, enhancing the effectiveness of wireless channels for secure communications regardless of the number, location, and channel conditions of the Tx, Rx, and malicious nodes. For example, if the nodes are less than a half-wavelength apart or the propagation environment has poor scattering, independent observations cannot be assumed owing to correlated channels. In such cases, a secure technique alone may not protect against all types of attacks. Therefore, a strategy involving the combination of different RIS-aided PLS solutions can provide high-level security enhancements \cite{m18,m19,m20}.

\begin{figure*}[t!]
\centering
\includegraphics[width=7in,height=6.8in]{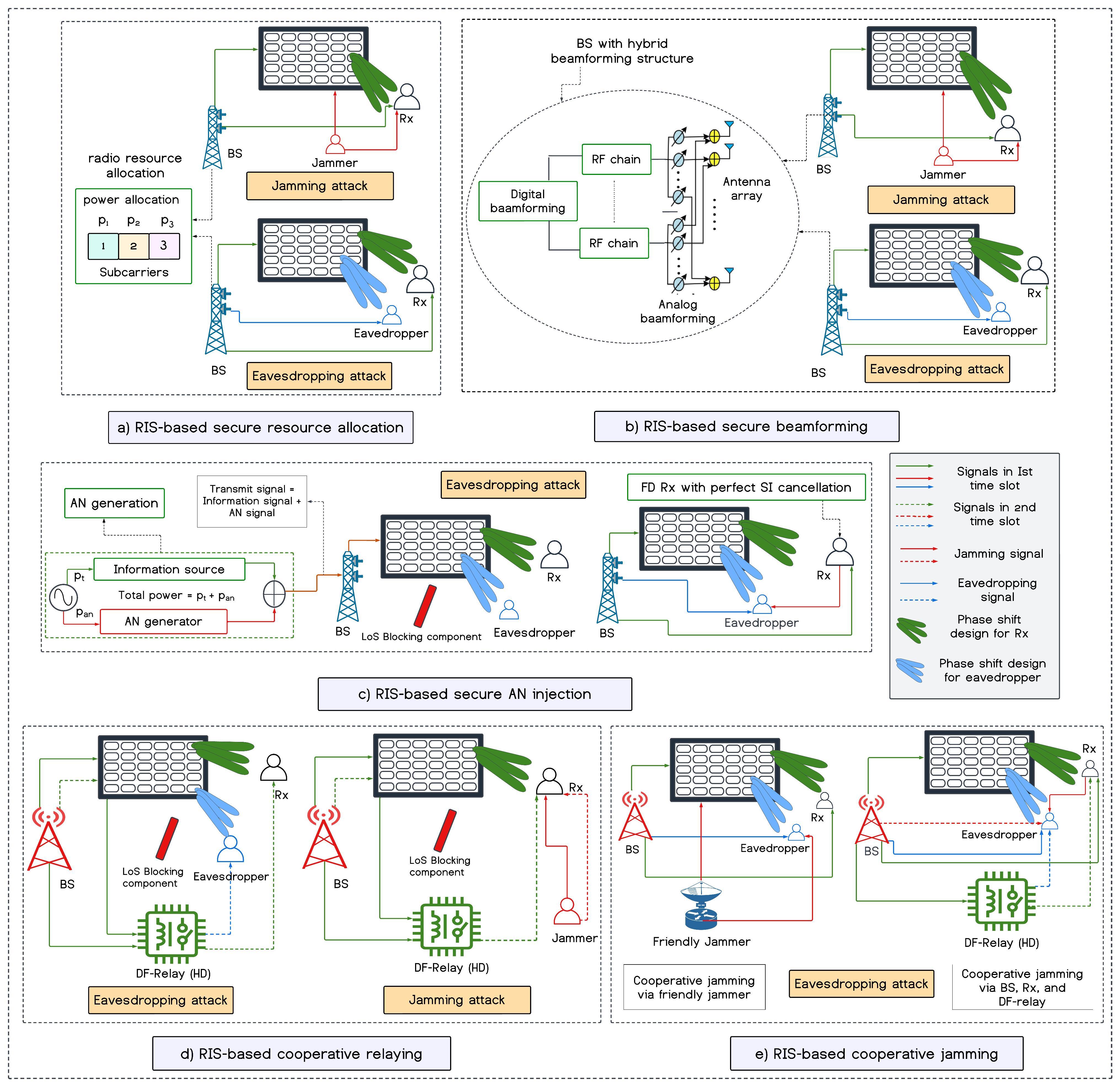}
\caption{Illustration of RIS-aided PLS design solutions: a) RIS-based secure resource allocation, b) RIS-based secure beamforming, c) RIS-based secure AN injection, d) RIS-based secure cooperative relaying, and e) RIS-based secure cooperative jamming.}
\label{fig4}
\end{figure*}

\subsubsection{RIS-Based Secure Resource Management} The effective utilization of network resources (bandwidth, time, and power) can ensure secure transmissions. As an example, link adaptation based on sub-carrier and power allocation, adaptive modulation, and coding can achieve the desired security objectives. PLS provision can be enabled by utilizing transmission parameters optimized specifically for the Rx link using independent fading. Because the signal intended for the Rx is optimized, it inherently provides security against jamming and eavesdropping attacks without requiring any additional processing or computation at the Rx \cite{m21,m22}. Furthermore, a more secure communication link can be established by adapting the link between the Tx and Rx via RIS (e.g., by adjusting the transmission parameters and the reflection characteristics). As illustrated in Fig. \ref{fig4}(a), the RIS can steer the transmitted signals towards the Rx while simultaneously canceling out the interference from the jammer and/or degrading the SNR at the eavesdropper. Thus, the RIS can effectively create a "secure zone" around the Tx and Rx, making it much more difficult for an eavesdropper (or a jammer) to intercept (or disrupt) the information. Specifically, RIS-based link adaptation and channel-dependent resource allocation can support flexible and scenario-specific secure transmissions and satisfy information security requirements in terms of confidentiality and availability. In this regard, certain parameters can be adjusted based on channel characteristics, such as transmit power, modulation order, number of RIS elements and their reflection coefficients, error correction, sub-carriers, coding rate, and RF bandwidth.

\subsubsection{RIS-Based Secure Beamforming}
The spatial dimension of multiple antennas in MIMO systems introduces signaling degrees of freedom that can be exploited to ensure the reliable transmission of information. Through beamforming and precoding techniques, one or more spatially directed signals are transmitted using an antenna array to achieve diversity and array gains. Beamforming involves directing EM energy toward a particular location by feeding the same signal to each antenna in the array and steering the signal using phase shifters to adjust the beamforming weights in real time. Precoding involves simultaneously transmitting multiple symbols (i.e., rank-k transmission) and shaping the radiation pattern of the antenna array at the Tx towards the Rx. In general, beamforming refers to a single-user transmission with one data stream, and precoding refers to the superposition of different beams for spatial multiplexing of multiple data streams. Examples of linear precoding for multi-user MIMO systems include generalized singular value decomposition, zero-force precoding, and dirty-paper precoding \cite{m23}. These signal processing techniques can also boost PLS performance by controlling the directivity or shape of the transmitted signals. For example, an eavesdropper must be equipped with the same or better MIMO configurations to intercept information transmitted over multiple spatial channels. The passive beamforming of the RIS (the phase-shift design of the elements) can complement the active beamforming of the Tx in providing PLS performance. As shown in Fig. \ref{fig4}(b), the optimal beamforming vectors can be designed to degrade the eavesdropping channel (relative to the legitimate channel) to achieve confidentiality and enhance the power of the decoding signal at the Rx while negating the impact of the jamming signal to achieve availability. This can increase the difficulty for the eavesdropper and/or jammer to intercept and/or disrupt information, respectively. Secure transmissions can be more efficient by increasing the number of elements of the RIS, compared to increasing the size of the antenna array of the Tx. For example, using fewer antennas at the Tx can provide a certain degree of secrecy gain when the RIS is deployed \cite{m24}. The higher the number of elements at the RIS, the lower the number of antennas required at the Tx. RIS-based secure beamforming designs can improve PLS performance metrics in different system configurations (e.g., multi-stream, multi-user, and wide-band). However, the use of a particular technique depends on the rank and security requirements of the transmissions.

\subsubsection{RIS-Based Secure AN Injection}
To prevent eavesdropping attacks, high-powered AN can be transmitted by a Tx or Rx. Specifically, the Tx can transmit AN in the same frequency band as the original signal by utilizing the null space in the MIMO channel between the Tx and Rx, while the Rx can generate AN by adopting in-band FD communications. For example, \cite{m25} deliberately introduced noise via the Rx by assuming noise-loop modulation. Furthermore, \cite{m26} considered AN precoding, i.e., a known interfering signal is added to the original signal at a Tx. Isotropic AN can be adopted in scenarios where the channel state information (CSI) of the eavesdropper is unknown. Otherwise, spatially selective AN can be adopted. Most studies on AN-based PLS assume knowledge of the location of the eavesdropper, which is not practicable. In a more realistic approach, the AN can be designed solely based on the information of the Rx. The signal can, however, still be intercepted by an attacker with sufficient advanced technology or channel knowledge. As shown in Fig. \ref{fig4}(c), the RIS can potentially reflect the AN and provide enough interference for the eavesdropper to make intercepting the signal more difficult. Additionally, in \cite{m26}, the authors demonstrated the improvement in secrecy provided by Tx-injected AN when the RIS is placed near the eavesdropper. It has been observed in \cite{m27} that the RIS with AN design required fewer elements (and less computation complexity) to achieve a certain level of secrecy compared to the design without AN. In addition, AN is typically a power-intensive approach because more transmission and circuit power are consumed. A constrained transmit power is required to transmit a confidential message to the Rx while injecting AN simultaneously, which can disrupt the decoding ability of the Rx. By utilizing the RIS, the Tx can potentially reduce the transmit power constraint, while maintaining high communication performance. The RIS-based AN injection optimization process generally involves selecting the optimal beams at the Tx/Rx (for transmitting/receiving the original signal and injecting the AN) using real-time CSI and designing phase shifts for the RIS.

\subsubsection{RIS-Based Secure Cooperative Communications} 
Cooperative relaying and jamming communications can utilize spatial diversity and provide efficient security solutions \cite{m28}. In cooperative relaying, multiple trusted relays ameliorate the PLS by providing the cooperative diversity benefits of MIMO technology in a distributed manner. To enhance reception performance at the Rx against eavesdropping and jamming attacks, relays can be selected based on their geographical locations or by using directional antennas to direct a narrow beam of radiation toward the Rx. In relay-based networks, the communication process is generally divided into two orthogonal phases, in which the first phase involves the transmission from the Tx (to both the relay and Rx) and the second phase involves the relay re-transmitting the signal to the Rx. The AF and DF protocols are typically used for relaying. The AF protocol is simpler to implement but suffers from noise amplification. The DF protocol performs better when the relay is positioned near the Tx and/or has better channel conditions. In accordance with transmission and reception capabilities, relays can be HD or FD. The advantages of relays have been extensively analyzed in challenging cases, such as unavailable or expensive backhaul methods, site acquisition problems for BS deployment, or high throughput requirements for cell edge or moving users. Cooperative relaying presents several open issues, including the selection of relays (which implies a trade-off between diversity benefit and security gain), reliability, power control, placement, hardware imperfections, and computational complexity. In addition to cooperative diversity, jamming signals can also be transmitted through relays in a coordinated manner to disrupt the interception of the eavesdropper.  The open issues for cooperative jamming include incentive mechanisms, power allocation under imperfect CSI, and the design of jamming signals against multiple eavesdroppers. Furthermore, hybrid cooperative relaying and jamming solutions can be used for PLS provision. By controlling radio propagation via RIS as shown in Figs. \ref{fig4}(d) and \ref{fig4}(e), cooperative diversity benefits can be further enhanced to improve PLS performance \cite{m29}.\\
$\underline{Remark}$: \textit{The secrecy performance improvement achieved by RIS-aided PLS designs can be a challenging problem owing to the high dimensionality of the channel space, high-precision configuration of several elements, strategic placement of the RIS, the coupled correlation of different parameters, and the stringent hardware constraints of the RIS. The other critical issues include the design of robust algorithms to adapt to external factors such as interference and noise, as well as incorporate the essential factors such as selection and/or scheduling of the relays in cooperative communications, power allocation in AN, antenna selection in beamforming, and the details of the RIS (e.g., geometry, orientation, surface material properties, cost, and implementation complexity), and the environment (e.g., channel conditions, signal frequency, available resources, and malicious nodes capabilities). The appropriate RIS-aided PLS solution can be selected based on design scenarios and communication objectives,  facilitating the trade-off between security performance, hardware complexity, latency, and overhead requirements.}

\begin{figure*}[t!]
\centering
\includegraphics[width=5.9in]{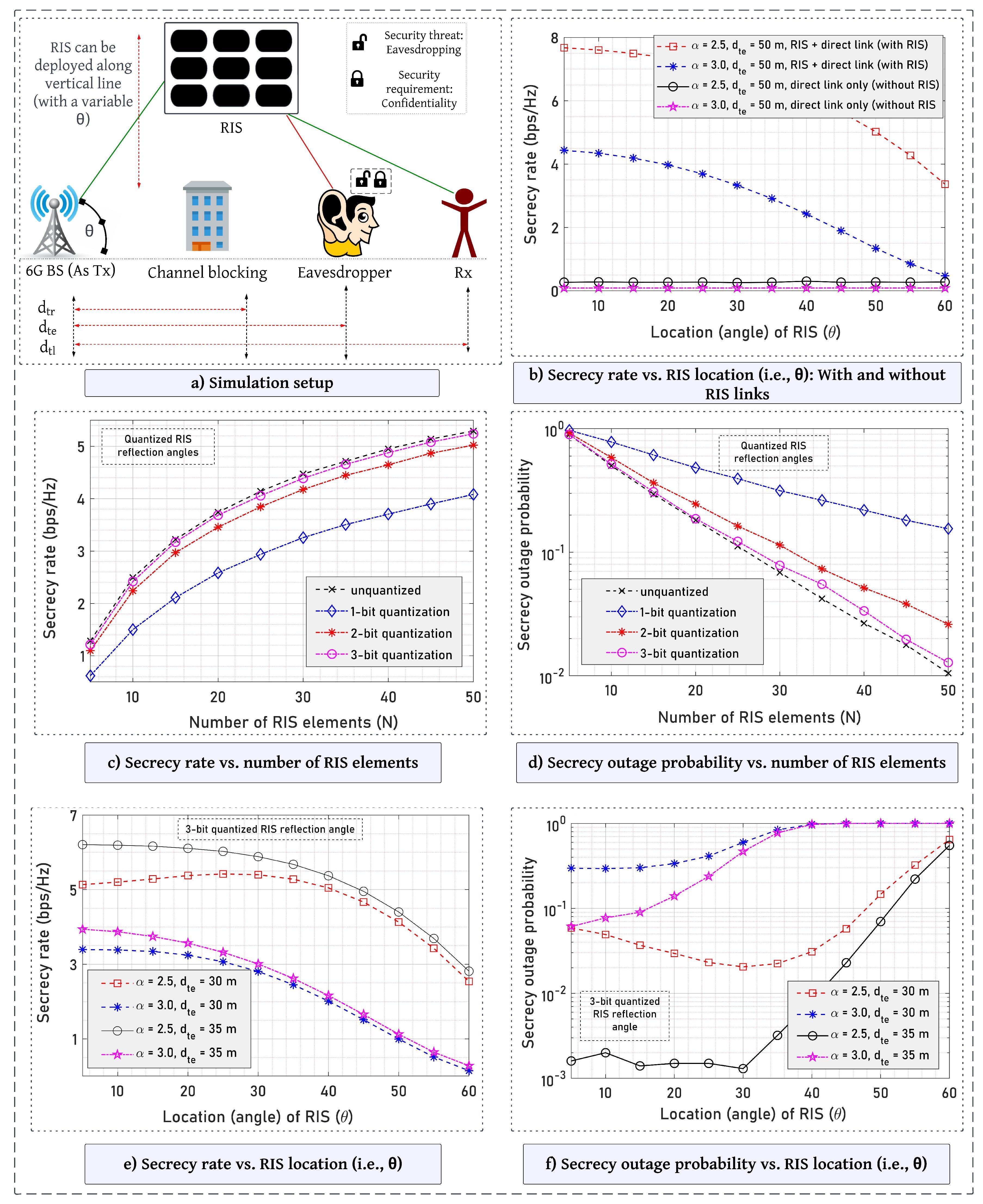}
\caption{Simulation results: a) Network topology, b) Secrecy rate vs. RIS location: with and without RIS links, c) Secrecy rate vs. Number of RIS elements, d) Secrecy outage probability vs. Number of RIS elements, e) Secrecy rate vs. RIS location, and f) Secrecy outage probability vs. RIS location.}
\label{fig5}
\end{figure*}

\subsection{Simulation Results}
In this subsection, simulation results are presented to demonstrate the effectiveness of the RIS against eavesdropping attacks. The network topology including a Tx with a single antenna, an RIS with $N$ number of reflecting elements, a (legitimate) Rx with a single antenna, and an eavesdropper with a single antenna is illustrated in Fig. \ref{fig5}(a). Here, $d_{tr}$ is the distance between a Tx and an RIS, $d_{te}$ represents the distance between a Tx and an eavesdropper, and $d_{tl}$ is the distance between a Tx and an Rx. To prevent channel blocking, the RIS can be positioned at an angle from the horizontal line connecting the Tx, the eavesdropper, and the legitimate Rx. In this case, if we consider the location of the Tx as an origin, that is, $(0, 0)$, the coordinates of the RIS are given by $\left(d_{tr}, d_{tr}\tan(\theta)\right)$, and the coordinates of the eavesdropper and the Rx are given by $(d_{te}, 0)$ and $(d_{tl},0)$, respectively. Furthermore, path loss and Rayleigh distributions characterize large-scale and small-scale fading characteristics, respectively. Specifically, the path loss is determined by
\begin{equation}
L(d)=C_0\left(\frac{d}{d_0}\right)^{-\alpha}   
\end{equation}
where $C_0=30$dB, $d_0=1$m, and path loss exponent $(\alpha)=\{2.5, 3.0\}$. Unless otherwise stated, the parameters are set as transmit power $=20$dBm, noise variance $=-100$dBm, number of reflecting elements ($N$) $=50$, and blocking loss in the direct links $=50$dB. If the channel coefficient between the Tx and the $i$th RIS element is $h_i$, and the channel coefficient between the $i$th RIS element and the Rx is $g_i$, then the optimal phase to maximize the signal strength at the Rx is given by $-\arg(h_i g_i)$. With this phase shift, all signal components from the RIS reflecting elements will have the same phase, thereby maximizing legitimate SNR. In terms of PLS, this phase design is the optimal approach when the eavesdropper's channel is unavailable. In practice, continuous phase shift values cannot be delivered from the Tx to the RIS. It is therefore assumed that the quantized phase information is delivered to the RIS via a control channel with limited bandwidth. In the simulations, 1, 2, and 3-bit quantization schemes are employed. For example, 3-bit quantization chooses the nearest phase among $\{0, \pm\frac{\pi}{4}, \pm\frac{\pi}{2}, \pm \frac{3\pi}{4}, \pi\}$. Moreover, we assume that $h_i$ and $g_i$ for all $i \in \{1, \cdots, N\}$ are available at the Tx.

Fig. \ref{fig5}(b) shows the improvement of the secrecy rate enabled by an RIS. The simulation parameters are set as $d_{tr}=20$m, $d_{te}=50$m, and $d_{tl}=40$m. The results present the secrecy rate versus the RIS location angle, $\theta$, and demonstrate the effectiveness of RIS through the comparison of two different cases of links, i.e., i) direct link only (without RIS) and ii) RIS + direct link (with RIS). In the results, we consider the unquantized RIS  phases. The higher $\alpha$ shows a lower secrecy rate because it degrades the quality of the legitimate channel more significantly than that of the eavesdropping channel. Moreover, as $\theta$ increases, the lower secrecy rate is achieved because the indirect path via RIS has a longer distance. In any case, we can achieve a much higher secrecy rate with RIS.

Figs. \ref{fig5}(c) and \ref{fig5}(d) show the secrecy rate and secrecy outage probability performance, respectively, depending on the number of RIS elements ($N$) with different quantization policies. Parameters are set as $d_{tr}=20$m, $d_{te}=30$m, $d_{tl}=40$m, $\theta=10^\circ$, and $\alpha=2.5$. As the number of RIS elements increases, the secrecy rate increases and the secrecy outage probability decreases. For secrecy outage probability, the threshold of the secrecy rate is set to be $2.5$ bps/Hz. The results also suggest that 3-bit quantization is sufficient to obtain the optimal secrecy performance provided by RIS in a given simulation scenario.

Finally, Figs. \ref{fig5}(e) and \ref{fig5}(f) illustrate the secrecy rate and secrecy outage probability, respectively, as a function of RIS location angle, i.e., $\theta$. Simulation parameters are set as $d_{tr}=20$m, $d_{te}=\{30, 35\}$m, $d_{tl}=40$m, and quantization bits for the RIS reflection angle $=3$. For secrecy outage probability, the threshold of the secrecy rate is set to be 3.0 bps/Hz. The results demonstrate that secrecy rates decrease and secrecy outage probabilities increase with $\theta$ as the distance to the RIS increases. Under a particular setup, an optimal $\theta$ maximizes the secrecy rate and minimizes the secrecy outage probability. For example, when $\alpha=2.5$ and $d_{te}=30$m, $\theta=25^\circ$ will achieve the highest secrecy rate and $\theta=30^\circ$ will achieve the lowest secrecy outage probability. Furthermore, secrecy performances are more affected by the path loss exponent than by the location of an eavesdropper. Results also illustrate how secrecy performance is negatively and positively impacted by $\alpha$ and $d_{te}$, respectively.

\section{Research Issues and Possible Solutions}
\label{section4}
This section discusses the research issues and possible solutions for RIS-aided secure 6G-IoT systems to provide design guidelines and useful insights.
\subsection{RIS Modeling}
\subsubsection{Research Issues}
The development of unified, physics-compliant, and ubiquitous hardware models for RIS-aided PLS designs is an active research area. It is an intrinsically complex task to analyze the functions of the RIS in different hardware implementations (e.g., PIN diodes, antennas, and NTT Docomo smart glasses) and to examine how the arbitrary EM field interacts with the RIS \cite{m30}. There has been limited research on RIS modeling that incorporates important parameters such as the coupling of the elements (based on the geometry and periodicity of the RIS), impedance matching, hardware impairments (e.g., RF losses, and limited or low-resolution phase shifts), and scattering properties of the elements.
\subsubsection{Possible Solutions}
The phase shift, load impedance, and generalized sheet transition conditions models can characterize the reconfigurability of elements and analyze the boundary conditions of the EM field at RIS. Through the development of novel hardware designs and manufacturing solutions, RIS systems can be made more scalable and affordable while maintaining tuning capabilities and continuous control. A quasi-continuous quantization method can be applied to amplitude and phase controls or to phase-only controls, with an inherent trade-off between implementation complexity and performance improvement.
\subsection{Channel Modeling}
\subsubsection{Research Issues}
A channel model integrates hardware design with communication theory and describes the complicated interaction between wireless signals and the radio environment mathematically. Path loss and multi-path fading are typically used to describe large-scale and small-scale channel characteristics, respectively. More research is required to develop numerically reproducible channel models for RIS-assisted PLS in 6G-IoT systems \cite{m31}.
\subsubsection{Possible Solutions}
A general standard channel modeling framework is required to accommodate complex propagation phenomena associated with different application scenarios. It may include large path loss and sparse multi-path components in mmWave/sub-THz, non-stationary and channel hardening in massive MIMO, large channel space and cascaded fading in RIS, high transmission loss underwater, random fluctuations and non-line-of-sight propagation in industrial IoT, large elevation and high mobility in UAVs, and a large coverage distance and delay spread in satellites \cite{m32}. 
\begin{itemize}
\item For accurate and tractable modeling, path loss models must incorporate the impact and scaling laws of design parameters (e.g., the size of the RIS, the placement of the RIS plane, the number of elements, and transmission distances). As 6G RISs have relatively large apertures and higher operating frequencies, near-field propagation, and indoor wireless solutions are essential.
\item There is a lack of physical implementation-compliant and mathematically traceable channel models required to evaluate the physical-layer performance under different deployment scenarios, channel conditions, antenna settings, network architectures, and system parameters. The channel models can be classified as deterministic (e.g., measurement-based, ray-tracing-based, map-based, and iterative-based) or stochastic (e.g., geometry-based and correlation-based). Owing to the complexity of exact fading distributions in complex environments, mathematical approximations are also critical \cite{m33}.
\end{itemize}
\subsection{Channel Estimation}
\subsubsection{Research Issues}
In RIS-aided PLS design, the performance gain depends on the reconfiguration of the RIS elements, which requires accurate, timely, and low-complex channel estimation. The general assumption of perfect CSI for both Rx and malicious users (eavesdropper and/or jammer) provides an upper bound on theoretical performance. It is challenging to acquire perfect CSI with negligible training overhead, computational complexity, and onboard signal processing. It is due to hardware constraints and nonlinear characteristics of RIS elements, a lack of information from malicious users (e.g., passive eavesdroppers are generally unknown and do not cooperate with the Tx), and errors in channel estimation and quantization that impair CSI estimation. The estimation of high-dimensional channels in RIS-aided secure 6G-IoT systems has been examined under various channel conditions and system scenarios \cite{m34}.
\subsubsection{Possible Solutions}
Passive elements lack RF chains to transmit or receive pilot signals to estimate unknown channel coefficients. To estimate individual channels via training signals, an RIS can be equipped with low-power RF chains. Otherwise, the concatenated channel can be estimated for a known RIS configuration by sharing a common channel coefficient for nearby elements with a trade-off in computational complexity and performance degradation. Thus, passive beamforming can be designed for secure transmissions based on feedback from the BS and users without explicitly estimating the individual channels. Pilot sequences and training patterns are necessary for channel estimation algorithms, such as least squares, minimum mean squared errors, and compressed sensing. The spatially sparse nature of mmWave and sub-THz channels can reduce training burden and performance error \cite{m35}. Nevertheless, accurate channel models and hardware designs are needed. Low-dimensional, quasi-static, and low-rank channel features can also be exploited to reduce pilot overhead. Channel estimation for secure 6G-IoT systems using RIS can aim at the following:
\begin{itemize}
\item Channel estimation methods with optimal performance and complexity trade-offs under different signal models and system configurations are crucial. Developing a training signal and estimation algorithm jointly is not trivial. Thus, optimizing a training signal for a particular estimation algorithm can be efficient.
\item  The challenges associated with channel estimation include phase noise, spatial correlation, absence of information from jammers and/or eavesdroppers, configuration of reflecting elements in real time, deployment of large-scale RIS at random locations, and propagation characteristics of mmWave and sub-THz signals.
\item  An imperfect CSI deteriorates security performance. Owing to this, it is imperative to design robust PLS designs under imperfect or error-prone CSI. However, uncertainty constraints associated with imperfect CSI models complicate optimization problems.
  \end{itemize}

\subsection{Optimization Approaches}
\subsubsection{Research Issues}
 The optimization of RIS-aided PLS designs constitutes a challenging mathematical problem owing to the multiple and coupled variables, which can be related to resource allocation (e.g., sub-carrier and power allocation), RIS (e.g., size, placement plane, and element number), beamforming (e.g., antenna selection), AN (e.g., power allocation factors), and cooperative communications (e.g., relay selection and/or placement). Non-convex objective functions (e.g., performance metrics, such as secrecy capacity, secrecy outage probability, and secure energy efficiency) and/or non-convex constraint functions (such as source power constraint, phase shift modulus constraint, and discrete constraint for RIS elements) make optimization problems non-convex and non-trivial. Fading channels usually require complex functions, making the analysis further complicated by the large channel space required for many users and various RIS elements. In addition, the complex topologies and nonlinear components in 6G increase the complexity of global optimization \cite{m36}. 
\subsubsection{Possible Solutions}
In RIS-aided secure 6G-IoT systems, optimization problems are diverse and complex, making it unacceptable to use exact and simple convex optimization methods, such as linear programming, dynamic programming, exhaustive search, and interior point analysis. Therefore, effective methods are needed to solve high-dimensional/multi-objective optimization problems. An analysis can be simplified by excluding hardware impairments and imperfect conditions, such as the phase-dependent amplitude of the RIS, discrete amplitude and phase adjustments of the RIS, transceiver hardware impairments, channel estimation error, and correlation among channels. Intractable and non-convex optimization problems can be solved analytically with approximation and relaxation techniques. Using iterative search algorithms or heuristic algorithms, a non-convex optimization problem can be transformed into approximate sub-problems under convex relaxation (using approximate sub-problems). It is possible to categorize optimization methods into integer programming (with integer decision variables), quadratic programming (with quadratic objective functions), mixed-integer programming (with discrete and continuous variables), a difference of convex functions programming (with objective functions as the subtraction of two convex functions), fractional programming (with the ratio of two nonlinear functions), and semidefinite programming (with the problems under linear equality and non-negative constraint functions). Non-convex optimization problems can also be handled iteratively using iterative algorithms and alternative optimization techniques, such as the penalty function method, dual decomposition, semidefinite relaxation, and successive convex approximation \cite{m37}.

\subsection{ML-Based approaches}
\subsubsection{Research Issues}
The analytical solutions rely on complex mathematical models that approximate the real environment and are only applicable under certain conditions owing to limited working flexibility. In addition, conventional solutions may not work for complex RIS-aided PLS designs because of RIS configurations, dynamic radio channels, and malicious user behavior. In contrast, ML-based solutions are ideal for designing and optimizing RIS-aided secure 6G-IoT systems, addressing Rx-side optimization, RIS-side passive beamforming, channel modeling and estimation, and combating jammers and eavesdroppers of unknown types \cite{m38}. 
\subsubsection{Possible Solutions}
Deep learning (DL) can improve channel estimation, tracking, and prediction under imperfect environments. Model-free mapping is possible with DL algorithms by using learnable parameters to alleviate uncertainties (e.g., rapidly changing scattering environments and hardware impairments) and yield flexible results. The DL algorithms can learn RIS channels efficiently in any system configuration and signal model and present better estimation performance under complex channel conditions (e.g., non-stationary and nonlinear temporal dynamics, and spatial correlations). Owing to its large search space and learning and mapping capabilities, deep reinforcement learning (DRL) can solve complex and high-dimensional optimization problems efficiently. Furthermore, DRL algorithms can handle problems with multiple (conflicting) objectives, such as joint optimization of coverage and secrecy capacity, and can control both continuous and discrete actions. However, a large channel space in RIS-aided secure 6G-IoT systems requires complicated training processes. Moreover, offline-trained models are more complex, consume more power, and do not match real-world conditions. Further research is required to develop learning algorithms that are less complex (e.g., with a lower training overhead) while maintaining a high level of accuracy and stability.

\begin{figure*}[t!]
\centering
\includegraphics[width=6.8in,height=4.6in]{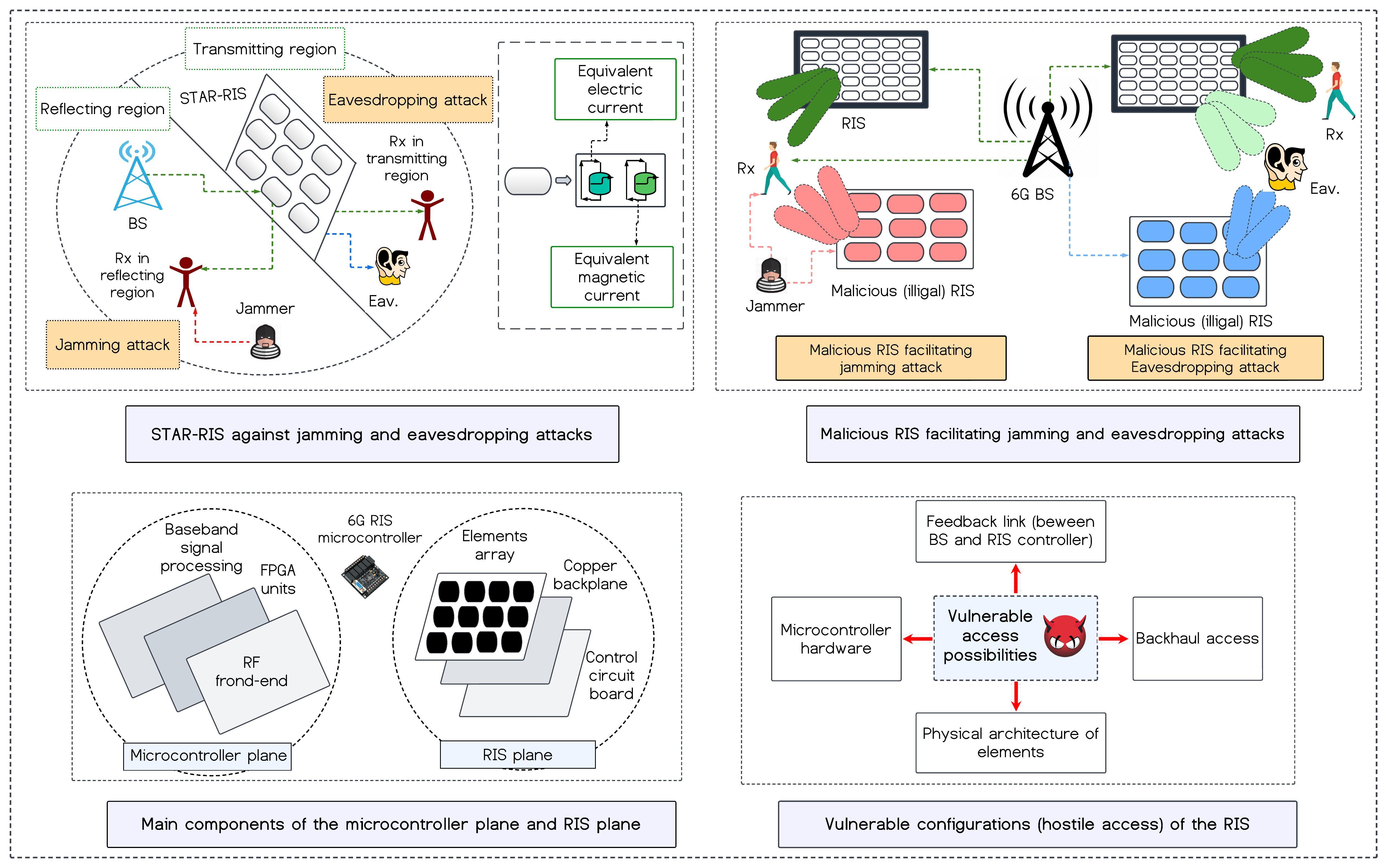}
\caption{An overview of STAR-RIS and malicious RIS for a secure 6G-IoT network.}
\label{fig6}
\end{figure*}

\section{Recent Advances}
\label{section5}
This section discusses the recent advances in terms of the STAR-RIS and malicious (illegal) RIS from the perspectives of eavesdropping and jamming attacks.
\subsection{STAR-RIS}
Reflecting-only RIS requires the Tx and Rx to be deployed on the same side. Therefore, service coverage cannot be provided for the Rx behind the RIS. STAR-RIS is a new paradigm for wireless communication that presents more degrees of freedom by manipulating the signals in a full space. STAR-RISs can provide both transmission (T) and reflection (R) signals to serve the users located on both sides \cite{m39}. From the hardware manufacturing perspective, STAR-RIS can enable electric and magnetic currents to produce T and R signals simultaneously or consecutively. Three operating protocols, energy splitting, mode selection, and time splitting, can achieve different communication modes through coupled or independent controls. Fig. \ref{fig6} demonstrates the effectiveness of the STAR-RIS against jamming and eavesdropping attacks. However, RIS-aided PLS designs cannot be directly applied to the STAR-RIS-aided PLS because STAR-RIS introduces adjustable parameters for both the T and R links. STAR-RIS-aided PLS designs can be directed as follows:
\begin{itemize}
\item The modeling and analysis of the STAR-RIS-aided PLS designs are still at an early stage. For example, the manipulation of both the T and R signals requires different hardware implementations (with different tuning mechanisms for the elements), channel modeling (to compute secure communications for the users in the T and R half-spaced regions), channel estimation (to design both T and R patterns), and its performance evaluation (e.g., pilot overhead, accuracy, and complexity depend on the number of users, elements, and operating protocols). 
\item The implementation of the STAR-RIS-aided PLS designs requires further research. Specifically, the beamforming and deployment of the STAR-RIS for the PLS in the presence of multiple users in a full space are far more complicated research problems. The selection of the appropriate STAR-RIS operating protocol based on the communication objective or a suitable balance between the design complexity and system performance is crucial.
\item The optimization problems of the STAR-RIS-aided PLS are complicated owing to the fundamental theoretical constraints of the STAR-RIS (e.g., coupled phase constraint in the ES protocol, unit modulus constraint in the TS and MS protocols), and constraints due to the location uncertainties of the eavesdropper and jammer. Specifically, beamforming designs with coupled T and R coefficients require novel convex and ML-based optimization algorithms owing to the requirement for a hybrid (i.e., continuous and discrete) control scheme. 
\end{itemize}

\subsection{Malicious RIS}
Rx and malicious nodes can exploit the RIS, such that both legitimate RIS and malicious (illegal) RIS can exist in the 6G-IoT wireless paradigm \cite{m40}. Fig. \ref{fig6} depicts the possible vulnerabilities of the configuration (access) of the RIS. The reasons may include its low hardware cost, imperfect reconfigurability of elements, and illegal deployment. Malicious RIS can facilitate jamming attacks by substantially improving the interference powers at the Rx or by extending the coverage of the interference signal. Effective reflection can cause deterioration (reduce the SINR at the Rx) by constituting (altering) the phase shift metric either by making it orthogonal (to the actual one) or by injecting the complex Gaussian distributed phase shifts. To facilitate an eavesdropping attack, malicious RIS can be deployed for more information leakage. This can be achieved by using an information-gathering approach (e.g., adopting CSI of the eavesdropper) to maximize the SNR of the eavesdropper instead of the Rx.

The adverse applications of the malicious RIS can be applied in both data transmission and pilot contamination. Malicious RIS can rapidly reshape illegitimate communication links and significantly increase the complexity of the PLS provision problem. For example, it is impossible to design the legitimate RIS to cancel the signaling of the malicious RIS owing to the covert deployment (and the absence of the CSI of the malicious RIS links). Furthermore, malicious RIS can considerably hinder pilot-based CSI methods and degrade the channel estimation accuracy of legitimate links. This disables the RIS-based channel estimation and beamforming schemes, making the optimization problems based on instantaneous CSI insufficient and outdated. Therefore, the adverse impact of the malicious RIS requires further analysis, and effective countermeasures are required to establish RIS-aided secure 6G-IoT networks. For example, optimization problems against jamming and eavesdropping attacks can be designed based on an imperfect CSI. In addition, case studies and performance evaluations can be conducted to validate the effectiveness of the RIS-aided PLS solutions in the presence of malicious RIS.

\section{Conclusion and Future Work}
\label{section6}

6G-IoT networks pose severe security risks owing to the broadcast transmission of confidential information and increased attack vectors. RIS and PLS can be combined to deliver lightweight security for IoT devices. This article analyzes RIS-aided PLS designs against eavesdropping and jamming attacks. The fundamental principles and hardware architecture of RIS and security concepts and methods of PLS are discussed. We present RIS-aided PLS design solutions based on resource allocation, beamforming, artificial noise, and cooperative communication. In addition, we discuss the research issues and potential solutions related to RIS modeling, channel modeling and estimation, optimization, and ML, as well as recent advancements, e.g., STAR-RIS and malicious RIS. In the future, the design solutions will be validated using hardware platforms and experiments and the research will be extended to practical settings, such as mmWave/THz propagation, RIS hardware limitations, eavesdroppers and jammers using ML and malicious RISs, and large-scale networks.

\section*{Acknowledgments}
\small{This research was supported by the Basic Science Research Program through the National Research Foundation of Korea (NRF) funded by the Ministry of Education(MOE) (NRF-2022R1I1A1A01071807, 2021R1I1A3041887), by the Institute of Information \& communications Technology Planning \& Evaluation (IITP) grant funded by the Korea government (Ministry of Science \& ICT (MSIT)) (2022-0-00704, Development of 3D-NET Core Technology for High-Mobility Vehicular Service), and by Korea University Grant.}

\end{document}